\documentclass{aastex63}
\usepackage{upgreek}
\usepackage[all]{hypcap}
\usepackage{amsmath}

\begin{document}

\title{
Observations of the Quiet Sun During the Deepest Solar Minimum of the Past Century with Chandrayaan-2 XSM -- Elemental Abundances in the Quiescent Corona}
    
\correspondingauthor{Santosh V. Vadawale}
\email{santoshv@prl.res.in}

\author[0000-0002-2050-0913]{Santosh V. Vadawale}
\affiliation{Physical Research Laboratory, Navrangpura, Ahmedabad, Gujarat-380 009, India }
\author[0000-0002-7020-2826]{Biswajit Mondal}
\affiliation{Physical Research Laboratory, Navrangpura, Ahmedabad, Gujarat-380 009, India }
\affiliation{Indian Institute of Technology Gandhinagar, Palaj, Gandhinagar, Gujarat-382 355, India}
\author[0000-0003-3431-6110]{N. P. S. Mithun}
\affiliation{Physical Research Laboratory, Navrangpura, Ahmedabad, Gujarat-380 009, India }
\affiliation{Indian Institute of Technology Gandhinagar, Palaj, Gandhinagar, Gujarat-382 355, India}
\author[0000-0002-4781-5798]{Aveek Sarkar}
\affiliation{Physical Research Laboratory, Navrangpura, Ahmedabad, Gujarat-380 009, India }
\author[0000-0003-2504-2576]{P. Janardhan}
\affiliation{Physical Research Laboratory, Navrangpura, Ahmedabad, Gujarat-380 009, India }
\author[0000-0001-5042-2170]{Bhuwan Joshi}
\affiliation{Physical Research Laboratory, Navrangpura, Ahmedabad, Gujarat-380 009, India }
\author[0000-0003-1693-453X]{Anil Bhardwaj}
\affiliation{Physical Research Laboratory, Navrangpura, Ahmedabad, Gujarat-380 009, India }
\author{M. Shanmugam}
\affiliation{Physical Research Laboratory, Navrangpura, Ahmedabad, Gujarat-380 009, India }
\author[0000-0002-0929-1401]{Arpit R. Patel}
\affiliation{Physical Research Laboratory, Navrangpura, Ahmedabad, Gujarat-380 009, India }
\author{Hitesh Kumar L. Adalja}
\affiliation{Physical Research Laboratory, Navrangpura, Ahmedabad, Gujarat-380 009, India }
\author[0000-0002-3153-537X]{Shiv Kumar Goyal}
\affiliation{Physical Research Laboratory, Navrangpura, Ahmedabad, Gujarat-380 009, India }
\author{Tinkal Ladiya}
\affiliation{Physical Research Laboratory, Navrangpura, Ahmedabad, Gujarat-380 009, India }
\author{Neeraj Kumar Tiwari}
\affiliation{Physical Research Laboratory, Navrangpura, Ahmedabad, Gujarat-380 009, India }
\author{Nishant Singh}
\affiliation{Physical Research Laboratory, Navrangpura, Ahmedabad, Gujarat-380 009, India }
\author{Sushil Kumar}
\affiliation{Physical Research Laboratory, Navrangpura, Ahmedabad, Gujarat-380 009, India }

\begin{abstract}

Elements with low First Ionization Potential (FIP)  are known to be three to four times more abundant in  active region loops of the solar corona than in the photosphere.
There have been observations suggesting that this observed ``FIP bias'' may be different in other parts of the solar corona and such 
observations are thus important in understanding the underlying mechanism.
The \textit{Solar X-ray Monitor} (XSM) on board the Chandrayaan-2 mission carried out spectroscopic observations of the Sun in soft X-rays during the 2019-20 solar minimum, considered to be the quietest solar minimum of the past century. These observations provided a unique opportunity to study soft X-ray spectra of the quiescent solar corona in the absence of any active regions. 
 By modelling high resolution broadband X-ray spectra from XSM, we estimate the temperature and emission measure
during periods of possibly the lowest solar X-ray intensity. We find that the derived parameters remain nearly constant over time with a temperature around ~2 MK, suggesting the emission is dominated by X-ray Bright Points (XBPs). We also obtain the abundances of Mg, Al, and Si relative to H, and find that the FIP bias is $\sim$2, lower than the values observed in active regions. 

\end{abstract}


\section{Introduction} \label{sec:intro}

Knowledge of elemental composition in the solar corona is crucial to understand various outstanding issues, such as energy / mass transfer between different atmospheric layers and the origin of the solar wind.  However, it is challenging to measure the absolute elemental abundances (i.e. relative to hydrogen) 
which leads to a common practice of measuring coronal elemental abundances relative to other elements.
One important problem related to the coronal elemental composition is the abundance enhancement of the low First Ionization Potential (FIP) elements (i.e. elements having FIP less than 10 eV), compared to their photospheric values, often termed as the FIP bias or FIP anomaly.
Observations of the FIP anomaly started with the pioneering work of \citet{Pottasch_63}.
Later, many researchers showed that the 
abundances of the low FIP elements in the corona can be as much as 3-4 times than that of the photosphere~\citep{Meyer_85,1992PhyS...46..202F,1999A&A...348..286F,2012ApJ...755...33S}. 
It was also observed that the FIP bias varies within different features of the corona \citep{Feldman_93} and shows variation with 
both the solar cycle and magnetic activity of the Sun~\citep{2017NatCo...8..183B,2018ARep...62..281P}.
A detailed review on the topic can be found in \cite{2018LRSP...15....5D}.

While the origin of the FIP bias is not fully understood, recent reports based on the EUV imaging spectroscopy (e.g., \citealp{2019A&A...624A..36D}, \citealp{2019ApJ...884..158D}) show that the low-temperature ($\sim$1 MK) non-active corona has nearly photospheric abundances. In contrast, hot loops (2-4 MK) at the core of the active region with high magnetic field show  stronger (3-4) FIP bias~\citep{1992PhyS...46..202F, Feldman_2000, Feldman_2003, Saba_1995, Zanna_2014}
Multiple theories have been proposed in literature to explain the FIP bias (see \citealp{2015LRSP...12....2L} for a review); however, the widely accepted theory is that based on the Ponderomotive force model~\citep{2004ApJ...614.1063L,2009ApJ...695..954L}.  This model can successfully explain the higher FIP bias in hot, magnetically closed loops as well as photospheric abundances in the relatively cooler open field structures.
It also predicts that higher magnetic activity may lead to higher FIP abundance in the solar corona.

Although early visible light solar eclipse observations~\citep{1975MNRAS.171..119M} measured coronal abundances relative to hydrogen,  most of the XUV spectroscopic observations determine abundances relative to some other elements, such as O or Si.
On the other hand, broad-band soft X-ray spectroscopic observations are capable of
measuring absolute abundances by considering the line to continuum ratio,
as initially proposed by \cite{1972SSRv...13..672W}
and attempted by \cite{1974ApJ...188..423W,
1974ApJ...192..169W}.
Lately, there have been multiple studies presenting measurement of absolute
abundances by self consistently modeling the continuum and characteristic
lines in the observed soft X-ray spectra~(e.g.
\citealp{2014SoPh..289.1585N,2014ApJ...786L...2W,2015ApJ...803...67D,2015ApJ...802L...2C,moore18,2020SoPh..295..175N,2020ApJ...904...20S}).
However, these reports are based on observations of solar flares or active regions, where the underlying continuum is easier to measure due to high X-ray flux.
Similar studies during quiet Sun periods have not been possible so far due
to very low signal as well as difficulties in measuring the real
continuum.

Here we present the first such study of quiet corona using Chandrayaan-2/Solar X-ray Monitor (XSM)~\citep{2014AdSpR..54.2021V,shanmugam20}
which observes the Sun as a star in the soft X-ray band.
These observations carried out during the 2019/20 solar minimum, believed to be the deepest minimum 
in the past hundred years~\citep{2011GeoRL..3820108J, 2015JGRA..120.5306J},
provided a unique opportunity for long duration solar X-ray observations in 
the absence of solar active regions, thereby enabling one to infer the temperature, emission measure and elemental 
abundances in the quiescent solar corona.
A companion paper (hereafter paper-II), presents a detailed investigation of the sub-A class microflares observed in the quiet Sun during 
this period.
 The rest of this article is organized as follows. Section 2 provides the details of observations and data analysis. Results are presented and discussed in Section 3 and finally summarized in Section 4.

\section{Observations and Data Analysis} \label{sec:obs}

Chandrayaan-2 XSM measures the disk integrated solar X-ray spectra in the energy range of 1 -- 15 keV~\citep{2014AdSpR..54.2021V,shanmugam20}. 
Its primary objective is to provide measurement of incident solar X-ray spectra on the Moon for estimation of elemental 
abundance on the lunar surface with remote fluorescence spectroscopy.
It has been designed to cover the wide intensity range of the solar X-rays all the way from
the quiet Sun to X-class solar flares.
XSM employs a Silicon Drift Detector (SDD) to measure the solar spectrum with an energy resolution better than 180 eV at 5.9 keV and a time
cadence of one second, which is the highest for
a broad-band solar X-ray spectrometer available so far~\citep{2020SoPh..295..139M}.

The visibility of the Sun varies with two pre-defined orbital seasons of 
the Chandrayaan-2 orbiter, namely `dawn-dusk' (DD) and `noon-midnight' (NM), 
arising because of the attitude configurations of the spacecraft in the 
lunar orbit and lasting for about three months each~\citep{vanitha20}.
The primary observing periods for XSM are the DD seasons, typically 
lasting from mid-February to mid-May and mid-August to 
mid-November~\citep{mithun20_gcal,2020SoPh..295..139M}. In the present 
work, we use the data from the first two DD seasons from September 12 
to November 20, 2019 (DD1) and February 14 to May 19, 2020 (DD2).

The XSM processing electronics generates X-ray spectrum on-board at every second. The raw (level-1) XSM data thus consists of one second spectra as well as other auxiliary information such as house-keeping parameters and observation geometry, organized as 
 day-wise FITS files. The standard level-2 calibrated data includes solar X-ray light curves in the full energy range of 1 -- 15 keV at one second cadence and full spectra at a cadence of 60 s. The XSM specific Data Analysis Software (XSMDAS)~\citep{mithun20_soft} is used for basic data reduction as well as for generating light curves and spectra with any user selected time bins greater than one second. 
The only other user input required for analysis is to select Good Time Intervals (GTI) for the generation of light curve and spectrum. The default GTI selection includes the conditions for nominal ranges of the instrument health parameters and excludes periods when the Sun angle is greater than $38^{\circ}$ or when the Sun is occulted by the Moon. It should be noted that the default condition on the Sun angle considers the radius of the Sun to be $3\nom{R}$ in order to avoid any partial exposure to the extended corona.

Since the XSM is fixed mounted on the Chandrayaan-2 spacecraft, the position 
of the Sun within it's FOV continuously changes throughout the orbit, 
resulting in a continuous change in the effective area of XSM. 
The XSMDAS provides two options to account for these variations: it can provide 
corrected count rate as if it were observed on-axis, typically used for light curves and time resolved spectra
saved as type-II PHA file;
or the spectra can be retained as observed counts and the variation of the
effective area are accumulated in a corresponding ancillary response file 
(ARF), typically used for time integrated spectra saved in type-I PHA file.

\begin{figure*}[h!]
\begin{center}
    \includegraphics[width=0.99\textwidth]{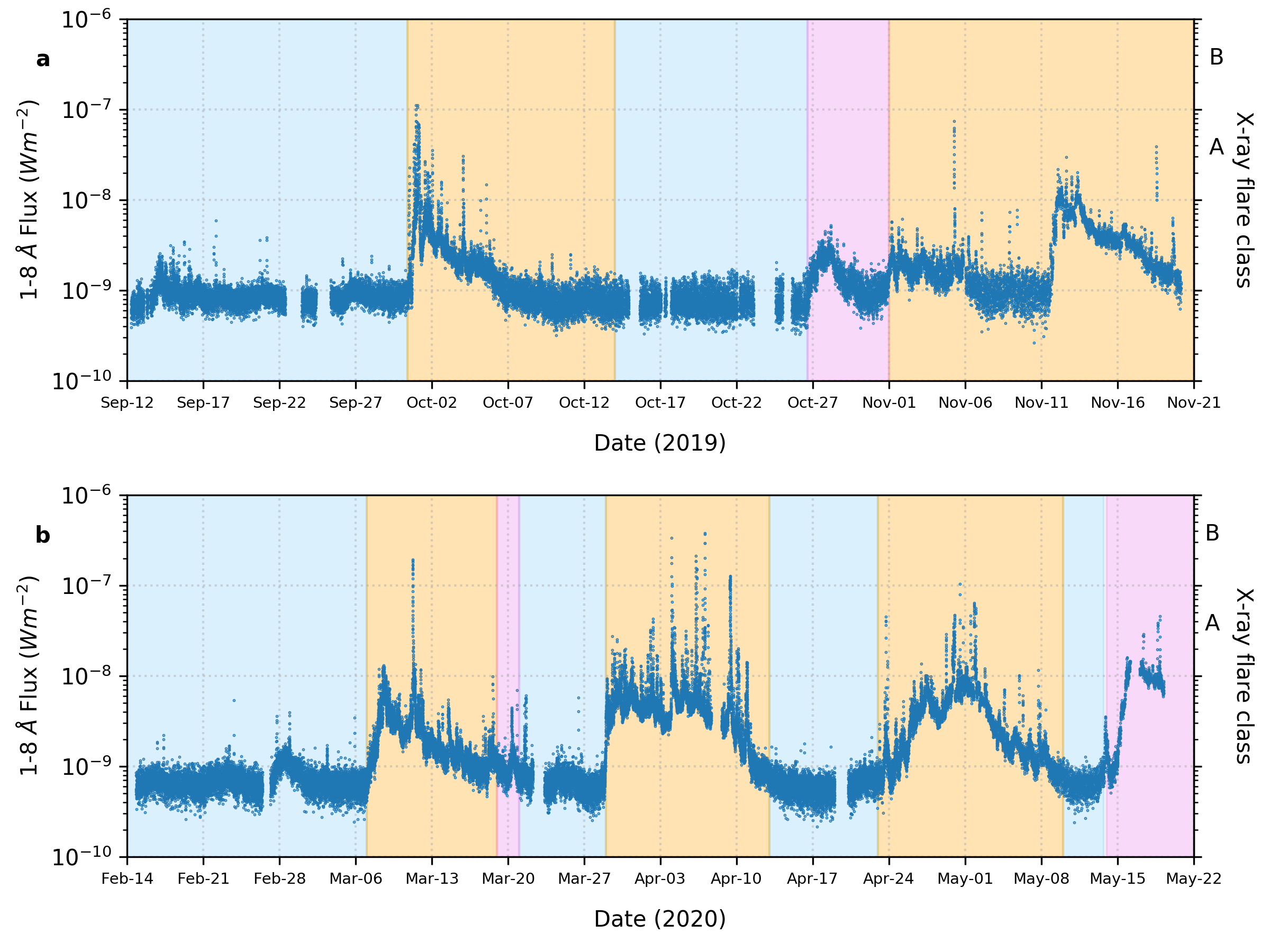}
    \caption{Solar X-ray flux in 1 -- 8 $\rm{\AA}$ (1.55 -- 12.4 keV) from XSM observations for the two DD seasons. Background colors 
    	in the figure correspond to intervals with active regions (orange), enhanced X-ray activity (pink), and quiet Sun 
    	observations (blue). 
    \label{xsmgoesrangeflux}}
\end{center}
\end{figure*}

For the present analysis, we use effective area corrected daily time-resolved spectra for obtaining flux light curves.
The time bin size was chosen to be two minutes so as to have sufficient counts in each spectra, given the very low X-ray 
intensity of the Sun.  The XSM flux light curve, $F(t)$ over any energy range $E_1$ to $E_2$  can then be generated from the 
type-II PHA files $S(E,t)$ using the equation:
\begin{equation}
F(t) = \sum_{E={E_1}}^{{E_2}} \frac{S(E,t)~E}{A(E)}
\end{equation} 

\noindent where $A(E)$ is the on-axis effective area of the XSM.  It should be noted that this assumes a diagonal redistribution 
matrix which, though not strictly correct, is adequate to estimate flux over broad energy ranges.
We then obtained the X-ray flux light curve in the energy range 1.55 -- 12.4 keV (same as the conventional 
GOES XRS band covering the wavelength range of 1 -- 8 $\rm{\AA}$) using time resolved spectra 
over the first two DD 
seasons, which is shown in Figure~\ref{xsmgoesrangeflux}.

\begin{figure*}[h!]
\begin{center}
    \includegraphics[width=0.99\textwidth]{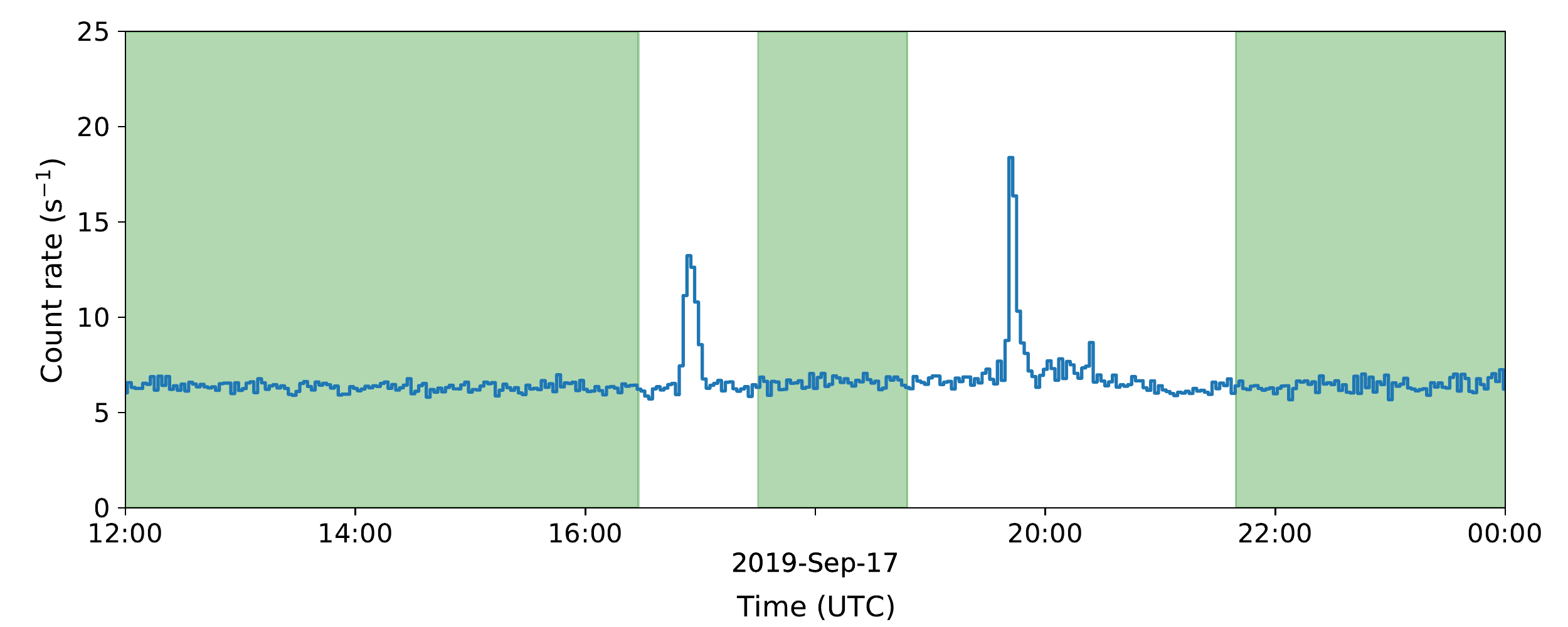}
    \caption{An example of selection of non-flaring quiescent period based on the XSM light curve is shown.     Durations
    	shaded green, which exclude the flare duration with additional margin before and after the flare,  
    	 were selected as periods of observation of the quiescent corona and used for 
    		spectral analysis.
    	}
    \label{QSGTI}
\end{center}
\end{figure*}

The flux light curve in Figure~\ref{xsmgoesrangeflux} clearly shows periods 
of elevated X-ray intensity due to the presence of active regions on the Sun. 
The orange background marks the periods when NOAA active regions 
were present on the solar disk. 
The pink background marks the periods when NOAA assigned active regions 
were not present, but the XSM light curve shows enhancement and corresponding
EUV and X-ray images from SDO/AIA and Hinode/XRT, respectively, show bright
regions.
Since the objective 
 of the present analysis was to focus on the quiet periods, we concentrated only on the periods 
 marked by the blue background in Figure~\ref{xsmgoesrangeflux} (a detailed analysis of the active region 
 emission will be presented elsewhere). The intervals selected for the present analysis are September 12-30 and October 14-26 in 2019 
 and February 14 - March 7, March 21-29, April 13-23, and May 10-13 in 2020, spanning a total of 76 days. We find that even during these intervals, when there were no active regions present on the Sun, 
 the XSM light curve shows a number of small flare-like episodes (microflares), which has been 
 discussed in paper-II. For the purpose of the spectroscopic investigation of the X-ray emission from a purely 
 quiescent corona, we conservatively ignore such microflares, along with 
 sufficient pre and post flare 
 buffer durations, obtained by visual inspection as shown in Figure~\ref{QSGTI}. These identified time intervals 
 were used as user GTI to generate quiet Sun spectra for carrying out a 
 detailed spectral analysis. 

To perform spectral fitting in XSPEC~\citep{arnaud96}, we generated XSM 
spectra (type-I PHA) and ARF for quiet Sun observations on each of the 
selected days with the user GTIs corresponding to the non-flaring periods.
Three days with very low exposures were ignored from further analysis. 
The non-solar background spectrum was obtained using XSM observations  
when the Sun was outside its FOV. Spectra below 1.3 keV were not used in fitting due to 
uncertainties in the response for the observations used in the present work~\citep{2020SoPh..295..139M}. 
For spectral fitting, we use an isothermal plasma emission model generated using the CHIANTI atomic database 
version 9.0.1~\citep{1997A&AS..125..149D,2019ApJS..241...22D}, which consists of the continuum and line emission. It is imported 
as a local model into XSPEC, with temperature, emission measure, and abundances of elements from Z=2 to 30 as model parameters. The details of the local model implementation will be discussed in a subsequent paper (Mondal et al., in prep.).

\section{Results and Discussion} \label{sec:result}

XSM observations during the first two DD seasons cover the period of possibly the lowest solar activity since the beginning of 
modern solar observations. The light curve shown in Figure~\ref{xsmgoesrangeflux} exhibits long periods when the solar X-ray 
intensity is very low but steady. It should be noted that the non-solar X-ray background measured by XSM  over the entire energy range is at least 35 times 
lower as discussed in ~\citet{2020SoPh..295..139M}. We find that the lowest solar X-ray flux measured by XSM in the GOES 1 -- 8 
$\rm{\AA}$ band is about $6\times10^{-10}~W~m^{-2}$, corresponding to the A0.06 class of solar activity, which is well below the 
sensitivity of the GOES-16 XRS instrument. 
Considering the fact that no active regions were present for an extended period during 
these observations, it is reasonable to assume that the solar corona was the quietest during these observations and that 
the XSM has measured the absolute floor level of the solar X-ray intensity.

\begin{figure}[h!]
\begin{center}
    \includegraphics[width=0.99\textwidth]{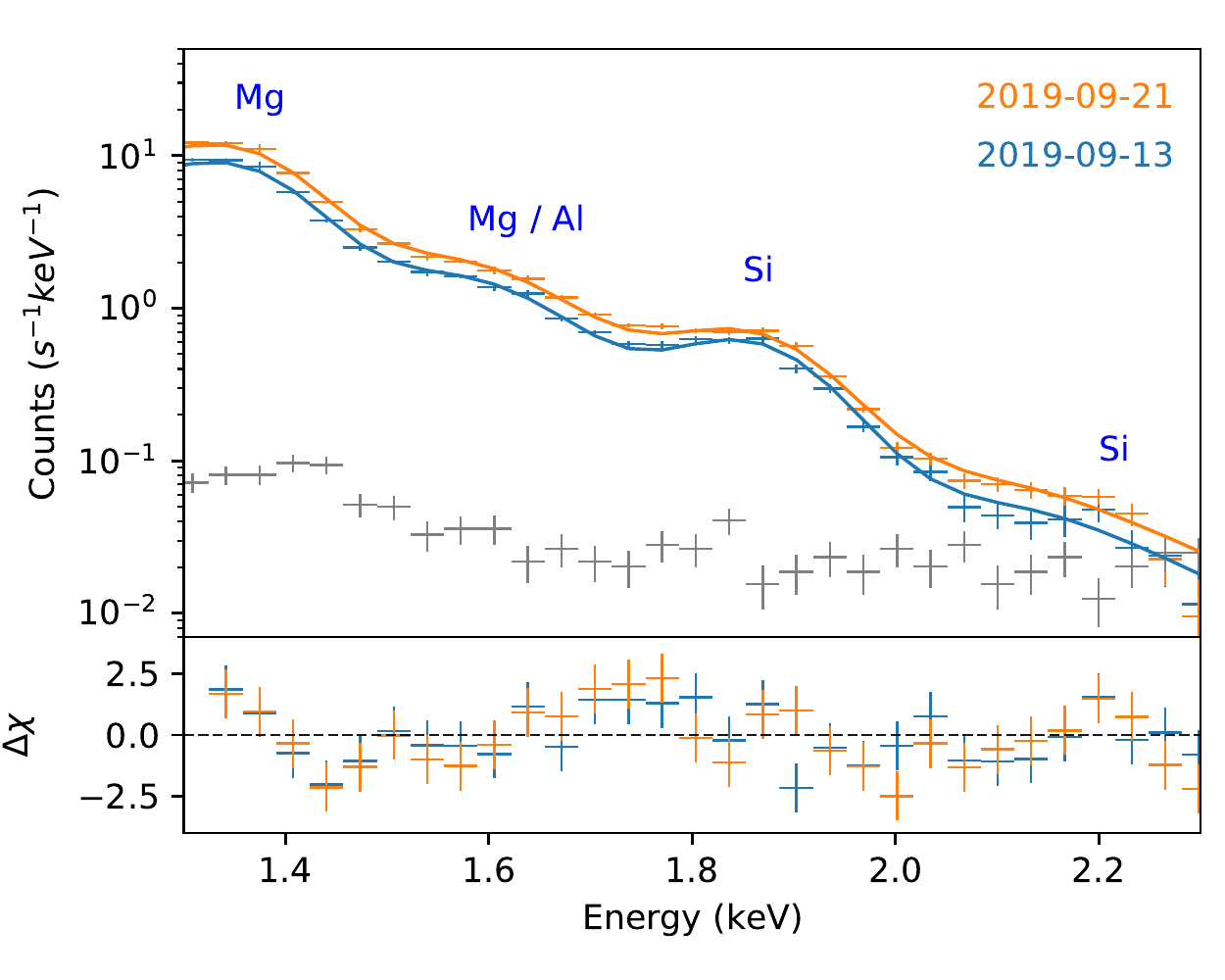}
    \caption{Soft X-ray spectra measured by the XSM for two representative days of quiet Sun observations are shown. Solid lines 
    	represent the best fit isothermal model and the residuals are shown in the bottom panel. Gray points correspond to 
    	non-solar background spectrum.}
        \label{xsmSpecFit}
\end{center}
\end{figure}

We find that the solar X-ray spectra integrated over any of the selected 73 days is dominant over the non-solar 
background spectrum up to 2.3 keV, as seen from Figure~\ref{xsmSpecFit}.
The spectra show a clear signature of thermal X-ray emission with the line complexes of Mg, Al, and Si. 
Hence, we fit the spectra in the energy range of 1.3 to 2.3 keV with the CHIANTI based isothermal plasma emission model that allows us to constrain the temperature, emission measure and abundances of Mg, Al, and Si. 
Abundances of all other elements, which do not contribute to the line emission in the energy range considered for fitting, are fixed to their known coronal abundance values. We verified that small changes in the abundances of these elements, or fixing them to their photospheric values, do not have any impact on the inferred
parameters. 
Figure~\ref{xsmSpecFit} shows the spectral fit results for two days of observation. It can be seen that the observed spectrum is well fitted with the isothermal model and similar fits were obtained for all spectra. One sigma errors on all free 
parameters of the model were also estimated using the standard procedure in XSPEC.

\begin{figure*}[h!]
\begin{center}
    \includegraphics[width=0.99\textwidth]{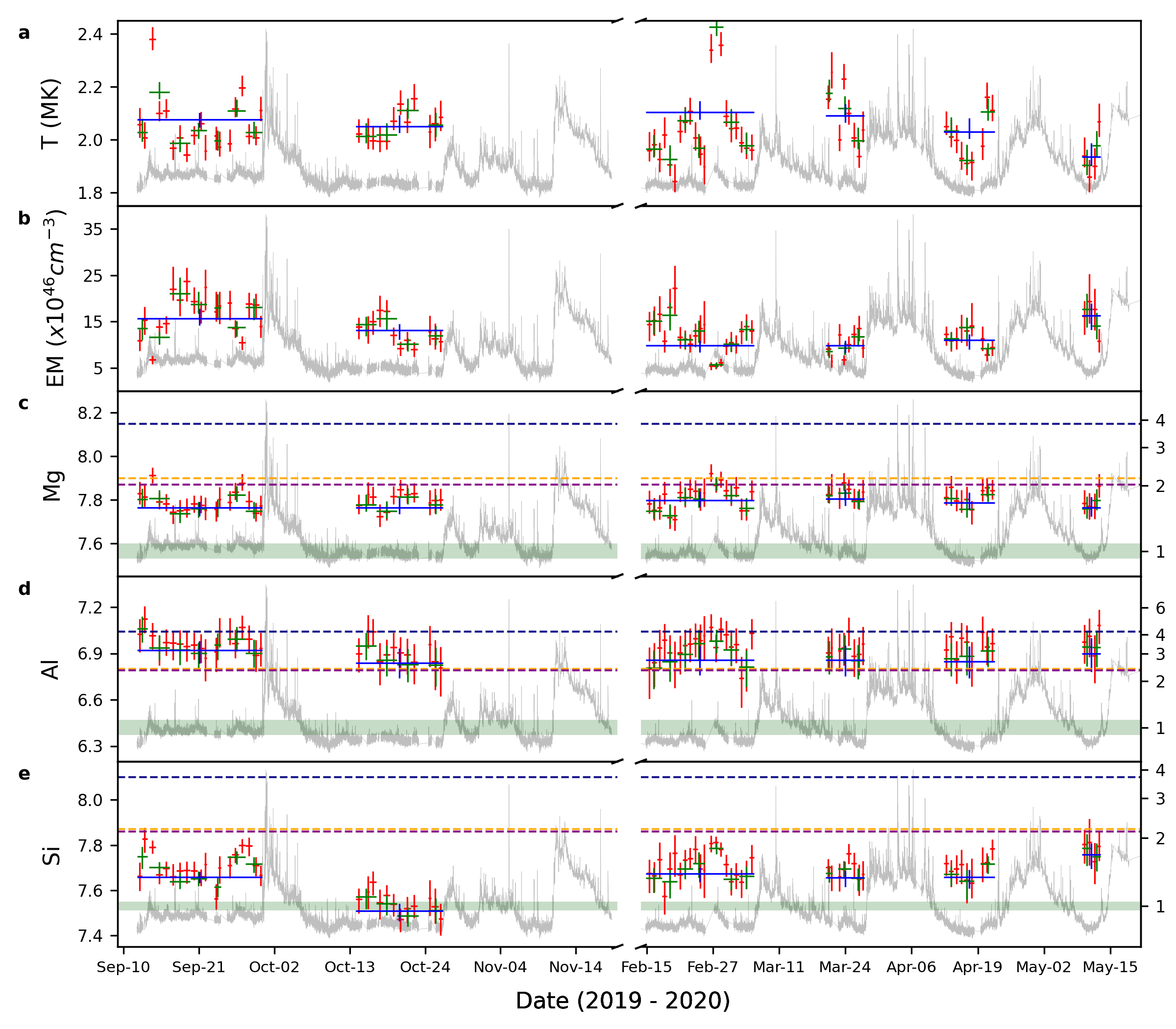}
    \caption{The five panels show the results of the spectral fitting, viz. Temperature
    ({\bf a}), EM ({\bf b}), as well as the absolute abundances of Mg ({\bf c}), Al ({\bf d}),
    and Si ({\bf e}) in logarithmic scale with A(H)=12. The red, green and blue points represent the best fit 
    parameters obtained from the spectra integrated over one day, multiple days (2--4 days), and 
    each quiet period, 
    respectively. The y-error bars represents 1$\sigma$ uncertainty for each parameter,
    whereas the x-error bars represent the duration over which a given spectrum is integrated.
    XSM light curves for the entire duration are shown in gray in the background.
    For a quick comparison with the reported values of abundances for these elements,
    the corresponding panels ({\bf c-e}) also show lines representing active region values reported by
    \cite{1992PhyS...46..202F} ({\em navy blue}), \cite{1999A&A...348..286F} ({\em orange}), and \cite{2012ApJ...755...33S} ({\em purple}). The range of photospheric abundances from various
    authors compiled in the CHIANTI database are shown as green bands.
    The right y-axis in panels {\bf c-e} show the FIP bias values for the respective elements with respect to average photospheric abundances.
\label{xsmQSAbund}}
\end{center}
\end{figure*}

By analysing integrated spectra for each day of the selected quiet Sun periods, we 
obtained temperature, EM, and abundances of Mg, Al, and Si as shown in Figure~\ref{xsmQSAbund}.
We find that the isothermal temperature and EM of the quiet corona typically remain constant
around $\sim$2.05 MK and ~$\sim1.5\times10^{47}~\rm{{cm}^{-3}}$, respectively.
 However, there are small variations in temperature and emission measure, which are correlated with the variations in X-ray flux.
\cite{2019SoPh..294..176S} reported isothermal temperatures of $\sim$1.69 MK for the quiescent corona using X-ray 
spectroscopic observations in a similar energy range using SphinX observations during 
the 2009 solar minimum, which is lower than the estimates from XSM. They also noted that the isothermal 
fit does not explain the observed spectra completely and had shown the presence of higher temperature 
components with DEM analysis, unlike in the present case where the XSM spectra in the range of 
1.3 -- 2.3 keV is consistent with isothermal models. 
One possible reason for the difference could be that the abundances were frozen to coronal values in the case of SphinX analysis as they could not be constrained due to relatively poorer energy resolution, whereas the abundances could be fitted in the case of XSM observations.

\begin{figure*}[h!]
\begin{center}
    \includegraphics[width=0.80\textwidth]{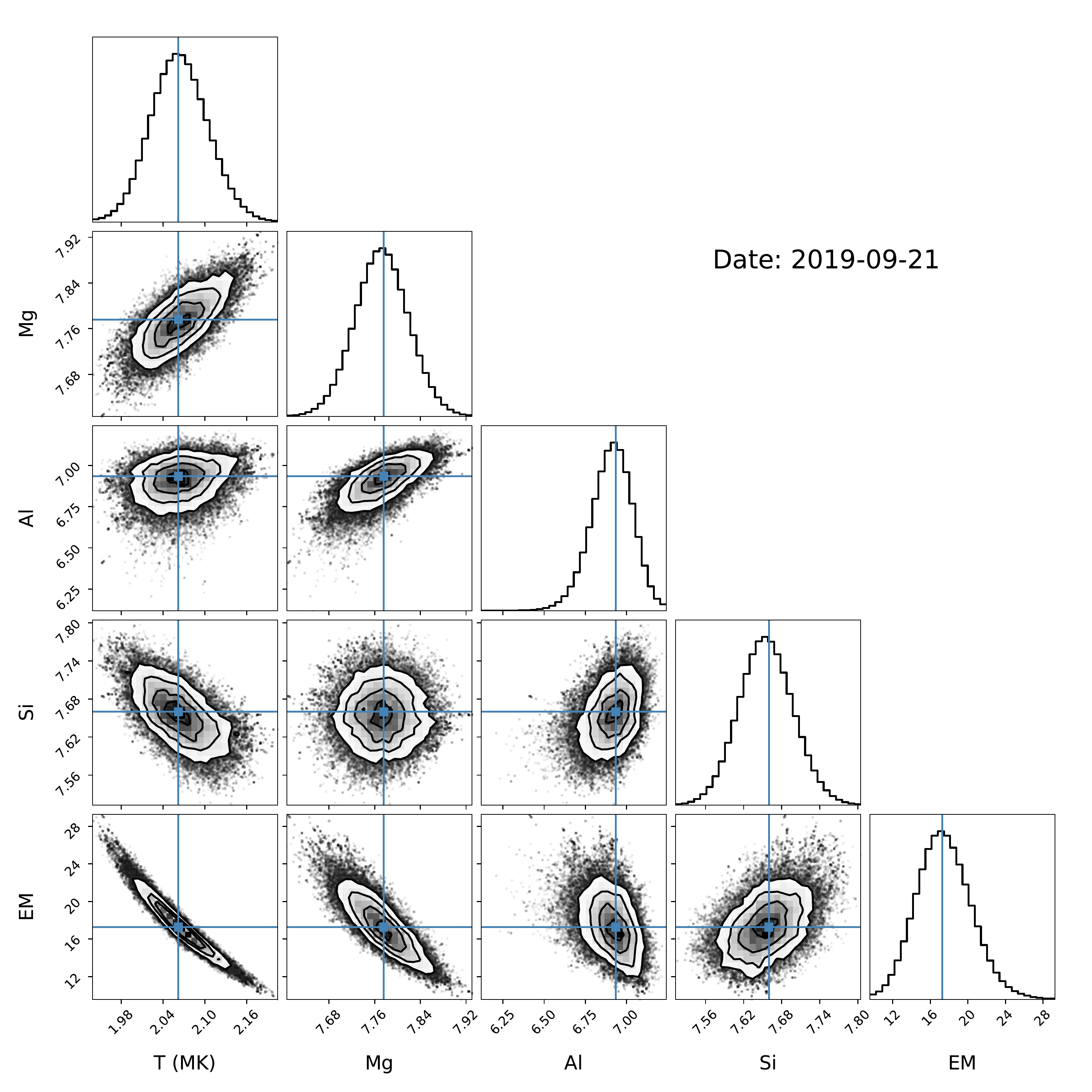}
    \caption{Corner plot showing results of MCMC analysis of representative quiet Sun spectrum on 21 September, 2019. The histograms represent marginalized distributions of each parameter. Correlations between all pairs of parameters are shown in the scatter plots overlaid by the contours corresponding to 1, 2, and 3 sigma levels. Best fit parameters are shown by the blue lines. Similar plots for all 73 days are available as a figure set in the online journal.}
        \label{mcmcResult}
\end{center}
\end{figure*}

The estimated abundances for the low FIP elements
Mg, Al, and Si are most of the time higher than the photospheric values.
However, compared to various coronal abundance values reported
in the literature for active regions~\citep{1992PhyS...46..202F, 1999A&A...348..286F,2012ApJ...755...33S},
our average values are 20 -- 60 \% lower for Mg and Si.
Whereas, for Al, the present derived values are $\sim$30 \% lower than
the ~\cite{1992PhyS...46..202F} value, but comparable with the others.
 We note that the contribution of Al in the energy band comprising of Al lines is about 10\%, resulting in a lower sensitivity to Al abundance as reflected in relatively larger error bars. In order to establish the robustness of the measurements of Al abundances as well as other parameters, we carried out Markov-Chain Monte-Carlo (MCMC) analysis and the results are shown in Figure~\ref{mcmcResult}. These results clearly show that all parameters including Al abundances are reasonably well constrained.
 To verify the consistency of our elemental abundance estimates over multiple days, we carried out analysis of the 
spectra integrated over 2-4 days. These results are also shown in the 
Figure~\ref{xsmQSAbund}.  We also carried out similar analysis for the spectra integrated over the whole duration 
of the respective quiet period (represented by blue lines in Figure~\ref{xsmQSAbund}) and the results are given in 
Table~\ref{qsfitpar}.  We note that the abundance of Si during the period of 14 to 26 October is anomalously low
compared to other selected periods and further investigations are needed to identify the reason behind this.

\begin{deluxetable}{c c c c c c}
\tablecaption{Quiet Sun parameters obtained from XSM spectra integrated over each quiet period.}
\label{qsfitpar}
\tablehead{
Observation & T & EM & Mg & Al & Si      \\
Period & (MK) & (10$^{46}$ cm$^{-3}$) &  &  &   
}
\startdata
2019 Sep 12 - Sep 30       &       $       2.07^{+0.02}_{-0.02}$&  $       15.6^{+2.08}_{-1.46}$&  $       7.76^{+0.02}_{-0.03}$&  $       6.92^{+0.06}_{-0.08}$&  $       7.65^{+0.02}_{-0.02}$\\
2019 Oct 14 - Oct 26       &       $       2.05^{+0.04}_{-0.02}$&  $       13.1^{+1.36}_{-2.04}$&  $       7.76^{+0.04}_{-0.02}$&  $       6.83^{+0.09}_{-0.09}$&  $       7.50^{+0.03}_{-0.04}$\\
2020 Feb 14 - Mar 7       &       $       2.10^{+0.04}_{-0.02}$&  $       9.86^{+1.07}_{-1.49}$&  $       7.79^{+0.04}_{-0.03}$&  $       6.85^{+0.09}_{-0.09}$&  $       7.67^{+0.02}_{-0.03}$\\
2020 Mar 21 - Mar 29       &       $       2.09^{+0.04}_{-0.02}$&  $       9.81^{+1.09}_{-1.54}$&  $       7.80^{+0.04}_{-0.03}$&  $       6.85^{+0.09}_{-0.10}$&  $       7.65^{+0.02}_{-0.04}$\\
2020 Apr 13 - Apr 23       &       $       2.02^{+0.05}_{-0.02}$&  $       10.9^{+1.24}_{-1.99}$&  $       7.78^{+0.04}_{-0.03}$&  $       6.84^{+0.09}_{-0.10}$&  $       7.65^{+0.03}_{-0.05}$\\
2020 May 10 - May 13       &       $       1.93^{+0.05}_{-0.03}$&  $       16.3^{+2.64}_{-3.13}$&  $       7.76^{+0.04}_{-0.04}$&  $       6.89^{+0.09}_{-0.11}$&  $       7.75^{+0.04}_{-0.06}$\\
\enddata
\end{deluxetable}

In order to investigate the reason for the different FIP bias in the XSM observations, we checked the 
X-ray images from the X-ray Telescope (XRT) on board Hinode~\citep{2007SoPh..243...63G} taken with the Be-thin filter, which has 
a similar efficiency as the XSM at lower energies. 
The X-ray images during the selected days show that most of the X-ray emission observed in the XSM energy range arises from 
a few hot spots, known as X-ray Bright Points (XBP), first reported by~\citet{1974ApJ...189L..93G}.
To verify this further, we simulated the expected XSM count rate from the quiet coronal region 
excluding any XBPs using the Differential Emission Measure (DEM) from \cite{2009ApJ...705.1522B}. 
We generated synthetic spectra with CHIANTI using this DEM and convolved it with the XSM detector response matrix 
to obtain the expected count rate.  We find that the X-ray emission from the diffuse corona having peak temperature 
around 1 MK and photospheric elemental abundances can account for only 30 to 50 \% of the observed count rate, suggesting 
that the majority of X-ray emission observed by XSM originates in XBPs. 
This is further confirmed by the fact that the overall temperature of around $\sim$2 MK, as observed by XSM, 
is much higher than that known for the quiet and diffuse corona. 
Observations with Hinode XRT have also reported temperatures ranging from 1.1 to 3.4 MK for XBPs~\citep{2011A&A...526A..78K}.
Thus, we conclude that the intermediate FIP bias observed by XSM most likely corresponds to the XBPs.
To the best of our knowledge, this is the first report of the elemental abundances for XBPs having a temperature of $\sim2$ MK.
According to present theoretical understanding of FIP bias~\citep{2009ApJ...695..954L,2016ApJ...831..160D},
the XBPs having intermediate field strengths and temperatures are expected to have intermediate FIP bias; however, there has been no observational
evidence so far supporting this conjecture. 
Our observations confirm this expectation for the first time
with robust abundance measurements over an extended period.

\section{Summary} \label{sec:summary}

In the Sun-as-a-star mode observations, carrying out prolonged study of the quiet solar corona is often challenging because of the 
presence of bright active regions that typically occur throughout the solar cycle. 
The 2019-20 solar minimum offered such an opportunity for extended quiet corona observations when there were no active regions 
present on the visible solar disk.
The XSM on board Chandrayaan-2 was the only X-ray spectrometer operational during a good part of this minimum and optimally 
utilized this opportunity. It measured possibly the lowest intensity of the coronal X-rays with high significance 
and we find that the bulk of this X-ray emission likely originates in the XBPs.
Detailed spectroscopic analysis of these observations show that plasma temperature in the XBPs is around $\sim2$ MK and 
that the abundances of the low FIP elements in the XBPs are at a level intermediate to their 
photospheric and coronal 
abundances. Our results are consistent with the ponderomotive force model which is 
widely considered to be responsible for the coronal FIP bias. 

Such a rare opportunity is not likely to be available at least for a decade, until the end of the solar cycle 25. 
Though the XSM may not be operational during next solar minimum, it will observe the Sun at least during the rising phase of 
Solar Cycle 25. Thus, with its superior sensitivity, energy resolution and time cadence, XSM is expected to provide rich observations 
having far reaching consequences for the study of highly dynamic Sun.

\acknowledgments
{
XSM was designed and developed by the Physical Research Laboratory (PRL), Ahmedabad
with support from the  Space Application Centre (SAC), Ahmedabad,
the U. R. Rao Satellite Centre (URSC), Bengaluru, and the Laboratory for Electro-Optics
Systems (LEOS), Bengaluru.
We thank various facilities and the technical teams of allthe above centers 
and Chandrayaan-2 project, mission operations, and ground segment teams for 
their support.
The Chandrayaan-2 mission is funded and managed by the Indian Space Research Organisation (ISRO).
Research at PRL is supported by the Department of Space, Govt. of India.
We gratefully acknowledge G. Del-Zanna, H. E. Mason, and U. Mitra Kraev for their help with the use of CHIANTI as well as very helpful discussions, facilitated through Royal Society Grant No. IES{\textbackslash}R2{\textbackslash}170199.
}
\vspace{5mm}
\facilities{Chandrayaan-2 (XSM)}
\software{XSMDAS~\citep{mithun20_soft}, XSPEC~\citep{arnaud96}, Python, Matplotlib,  Corner.py~\citep{2016JOSS....1...24F}}



\begin{thebibliography}{}
\expandafter\ifx\csname natexlab\endcsname\relax\def\natexlab#1{#1}\fi
\providecommand{\url}[1]{\href{#1}{#1}}
\providecommand{\dodoi}[1]{doi:~\href{http://doi.org/#1}{\nolinkurl{#1}}}
\providecommand{\doeprint}[1]{\href{http://ascl.net/#1}{\nolinkurl{http://ascl.net/#1}}}
\providecommand{\doarXiv}[1]{\href{https://arxiv.org/abs/#1}{\nolinkurl{https://arxiv.org/abs/#1}}}

\bibitem[{{Arnaud}(1996)}]{arnaud96}
{Arnaud}, K.~A. 1996, in Astronomical Society of the Pacific Conference Series,
  Vol. 101, Astronomical Data Analysis Software and Systems V, ed. G.~H.
  {Jacoby} \& J.~{Barnes}, 17

\bibitem[{{Brooks} {et~al.}(2017){Brooks}, {Baker}, {van Driel-Gesztelyi}, \&
  {Warren}}]{2017NatCo...8..183B}
{Brooks}, D.~H., {Baker}, D., {van Driel-Gesztelyi}, L., \& {Warren}, H.~P.
  2017, Nature Communications, 8, 183, \dodoi{10.1038/s41467-017-00328-7}

\bibitem[{{Brooks} {et~al.}(2009){Brooks}, {Warren}, {Williams}, \&
  {Watanabe}}]{2009ApJ...705.1522B}
{Brooks}, D.~H., {Warren}, H.~P., {Williams}, D.~R., \& {Watanabe}, T. 2009,
  \apj, 705, 1522, \dodoi{10.1088/0004-637X/705/2/1522}

\bibitem[{{Caspi} {et~al.}(2015){Caspi}, {Woods}, \&
  {Warren}}]{2015ApJ...802L...2C}
{Caspi}, A., {Woods}, T.~N., \& {Warren}, H.~P. 2015, \apjl, 802, L2,
  \dodoi{10.1088/2041-8205/802/1/L2}

\bibitem[{{Dahlburg} {et~al.}(2016){Dahlburg}, {Laming}, {Taylor}, \&
  {Obenschain}}]{2016ApJ...831..160D}
{Dahlburg}, R.~B., {Laming}, J.~M., {Taylor}, B.~D., \& {Obenschain}, K. 2016,
  \apj, 831, 160, \dodoi{10.3847/0004-637X/831/2/160}

\bibitem[{{Del Zanna}(2019)}]{2019A&A...624A..36D}
{Del Zanna}, G. 2019, \aap, 624, A36, \dodoi{10.1051/0004-6361/201834842}

\bibitem[{{Del Zanna} \& {Mason}(2014)}]{Zanna_2014}
{Del Zanna}, G., \& {Mason}, H.~E. 2014, \aap, 565, A14,
  \dodoi{10.1051/0004-6361/201423471}

\bibitem[{{Del Zanna} \& {Mason}(2018)}]{2018LRSP...15....5D}
---. 2018, Living Reviews in Solar Physics, 15, 5,
  \dodoi{10.1007/s41116-018-0015-3}

\bibitem[{{Dennis} {et~al.}(2015){Dennis}, {Phillips}, {Schwartz}, {Tolbert},
  {Starr}, \& {Nittler}}]{2015ApJ...803...67D}
{Dennis}, B.~R., {Phillips}, K. J.~H., {Schwartz}, R.~A., {et~al.} 2015, \apj,
  803, 67, \dodoi{10.1088/0004-637X/803/2/67}

\bibitem[{{Dere} {et~al.}(2019){Dere}, {Del Zanna}, {Young}, {Landi}, \&
  {Sutherland}}]{2019ApJS..241...22D}
{Dere}, K.~P., {Del Zanna}, G., {Young}, P.~R., {Landi}, E., \& {Sutherland},
  R.~S. 2019, \apjs, 241, 22, \dodoi{10.3847/1538-4365/ab05cf}

\bibitem[{{Dere} {et~al.}(1997){Dere}, {Landi}, {Mason}, {Monsignori Fossi}, \&
  {Young}}]{1997A&AS..125..149D}
{Dere}, K.~P., {Landi}, E., {Mason}, H.~E., {Monsignori Fossi}, B.~C., \&
  {Young}, P.~R. 1997, \aaps, 125, 149, \dodoi{10.1051/aas:1997368}

\bibitem[{{Doschek} \& {Warren}(2019)}]{2019ApJ...884..158D}
{Doschek}, G.~A., \& {Warren}, H.~P. 2019, \apj, 884, 158,
  \dodoi{10.3847/1538-4357/ab426e}

\bibitem[{{Feldman}(1992)}]{1992PhyS...46..202F}
{Feldman}, U. 1992, \physscr, 46, 202, \dodoi{10.1088/0031-8949/46/3/002}

\bibitem[{{Feldman} \& {Laming}(2000)}]{Feldman_2000}
{Feldman}, U., \& {Laming}, J.~M. 2000, \physscr, 61, 222,
  \dodoi{10.1238/Physica.Regular.061a00222}

\bibitem[{{Feldman} \& {Widing}(1993)}]{Feldman_93}
{Feldman}, U., \& {Widing}, K.~G. 1993, \apj, 414, 381, \dodoi{10.1086/173084}

\bibitem[{{Feldman} \& {Widing}(2003)}]{Feldman_2003}
---. 2003, \ssr, 107, 665, \dodoi{10.1023/A:1026103726147}

\bibitem[{{Fludra} \& {Schmelz}(1999)}]{1999A&A...348..286F}
{Fludra}, A., \& {Schmelz}, J.~T. 1999, \aap, 348, 286

\bibitem[{{Foreman-Mackey}(2016)}]{2016JOSS....1...24F}
{Foreman-Mackey}, D. 2016, The Journal of Open Source Software, 1, 24,
  \dodoi{10.21105/joss.00024}

\bibitem[{{Golub} {et~al.}(1974){Golub}, {Krieger}, {Silk}, {Timothy}, \&
  {Vaiana}}]{1974ApJ...189L..93G}
{Golub}, L., {Krieger}, A.~S., {Silk}, J.~K., {Timothy}, A.~F., \& {Vaiana},
  G.~S. 1974, \apjl, 189, L93, \dodoi{10.1086/181472}

\bibitem[{{Golub} {et~al.}(2007){Golub}, {Deluca}, {Austin}, {Bookbinder},
  {Caldwell}, {Cheimets}, {Cirtain}, {Cosmo}, {Reid}, {Sette}, {Weber},
  {Sakao}, {Kano}, {Shibasaki}, {Hara}, {Tsuneta}, {Kumagai}, {Tamura},
  {Shimojo}, {McCracken}, {Carpenter}, {Haight}, {Siler}, {Wright}, {Tucker},
  {Rutledge}, {Barbera}, {Peres}, \& {Varisco}}]{2007SoPh..243...63G}
{Golub}, L., {Deluca}, E., {Austin}, G., {et~al.} 2007, \solphys, 243, 63,
  \dodoi{10.1007/s11207-007-0182-1}

\bibitem[{{Janardhan} {et~al.}(2011){Janardhan}, {Bisoi}, {Ananthakrishnan},
  {Tokumaru}, \& {Fujiki}}]{2011GeoRL..3820108J}
{Janardhan}, P., {Bisoi}, S.~K., {Ananthakrishnan}, S., {Tokumaru}, M., \&
  {Fujiki}, K. 2011, \grl, 38, L20108, \dodoi{10.1029/2011GL049227}

\bibitem[{{Janardhan} {et~al.}(2015){Janardhan}, {Bisoi}, {Ananthakrishnan},
  {Tokumaru}, {Fujiki}, {Jose}, \& {Sridharan}}]{2015JGRA..120.5306J}
{Janardhan}, P., {Bisoi}, S.~K., {Ananthakrishnan}, S., {et~al.} 2015, Journal
  of Geophysical Research (Space Physics), 120, 5306,
  \dodoi{10.1002/2015JA021123}

\bibitem[{{Kariyappa} {et~al.}(2011){Kariyappa}, {Deluca}, {Saar}, {Golub},
  {Dam{\'e}}, {Pevtsov}, \& {Varghese}}]{2011A&A...526A..78K}
{Kariyappa}, R., {Deluca}, E.~E., {Saar}, S.~H., {et~al.} 2011, \aap, 526, A78,
  \dodoi{10.1051/0004-6361/201014878}

\bibitem[{{Laming}(2004)}]{2004ApJ...614.1063L}
{Laming}, J.~M. 2004, \apj, 614, 1063, \dodoi{10.1086/423780}

\bibitem[{{Laming}(2009)}]{2009ApJ...695..954L}
---. 2009, \apj, 695, 954, \dodoi{10.1088/0004-637X/695/2/954}

\bibitem[{{Laming}(2015)}]{2015LRSP...12....2L}
---. 2015, Living Reviews in Solar Physics, 12, 2, \dodoi{10.1007/lrsp-2015-2}

\bibitem[{{Mason}(1975)}]{1975MNRAS.171..119M}
{Mason}, H.~E. 1975, \mnras, 171, 119, \dodoi{10.1093/mnras/171.1.119}

\bibitem[{{Meyer}(1985)}]{Meyer_85}
{Meyer}, J.~P. 1985, \apjs, 57, 173, \dodoi{10.1086/191001}

\bibitem[{{Mithun} {et~al.}(2020){Mithun}, {Vadawale}, {Sarkar}, {Shanmugam},
  {Patel}, {Mondal}, {Joshi}, {Janardhan}, {Adalja}, {Goyal}, {Ladiya},
  {Tiwari}, {Singh}, {Kumar}, {Tiwari}, {Modi}, \&
  {Bhardwaj}}]{2020SoPh..295..139M}
{Mithun}, N.~P.~S., {Vadawale}, S.~V., {Sarkar}, A., {et~al.} 2020, \solphys,
  295, 139, \dodoi{10.1007/s11207-020-01712-1}

\bibitem[{{Mithun} {et~al.}(2021{\natexlab{a}}){Mithun}, {Vadawale},
  {Shanmugam}, {Patel}, {Tiwari}, {Adalja}, {Goyal}, {Ladiya}, {Singh},
  {Kumar}, {Tiwari}, {Modi}, {Mondal}, {Sarkar}, {Joshi}, {Janardhan}, \&
  {Bhardwaj}}]{mithun20_gcal}
{Mithun}, N.~P.~S., {Vadawale}, S.~V., {Shanmugam}, M., {et~al.}
  2021{\natexlab{a}}, Experimental Astronomy, 51, 33,
  \dodoi{10.1007/s10686-020-09686-5}

\bibitem[{{Mithun} {et~al.}(2021{\natexlab{b}}){Mithun}, {Vadawale}, {Patel},
  {Shanmugam}, {Chakrabarty}, {Konar}, {Sarvaiya}, {Padia}, {Sarkar}, {Kumar},
  {Jangid}, {Sarda}, {Shah}, \& {Bhardwaj}}]{mithun20_soft}
{Mithun}, N.~P.~S., {Vadawale}, S.~V., {Patel}, A.~R., {et~al.}
  2021{\natexlab{b}}, Astronomy and Computing, 34, 100449,
  \dodoi{https://doi.org/10.1016/j.ascom.2021.100449}

\bibitem[{{Moore} {et~al.}(2018){Moore}, {Caspi}, {Woods}, {Chamberlin},
  {Dennis}, {Jones}, {Mason}, {Schwartz}, \& {Tolbert}}]{moore18}
{Moore}, C.~S., {Caspi}, A., {Woods}, T.~N., {et~al.} 2018, \solphys, 293, 21,
  \dodoi{10.1007/s11207-018-1243-3}

\bibitem[{{Narendranath} {et~al.}(2014){Narendranath}, {Sreekumar}, {Alha},
  {Sankarasubramanian}, {Huovelin}, \& {Athiray}}]{2014SoPh..289.1585N}
{Narendranath}, S., {Sreekumar}, P., {Alha}, L., {et~al.} 2014, \solphys, 289,
  1585, \dodoi{10.1007/s11207-013-0410-9}

\bibitem[{{Narendranath} {et~al.}(2020){Narendranath}, {Sreekumar}, {Pillai},
  {Panini}, {Sankarasubramanian}, \& {Huovelin}}]{2020SoPh..295..175N}
{Narendranath}, S., {Sreekumar}, P., {Pillai}, N.~S., {et~al.} 2020, \solphys,
  295, 175, \dodoi{10.1007/s11207-020-01738-5}

\bibitem[{{Pipin} \& {Tomozov}(2018)}]{2018ARep...62..281P}
{Pipin}, V.~V., \& {Tomozov}, V.~M. 2018, Astronomy Reports, 62, 281,
  \dodoi{10.1134/S1063772918040054}

\bibitem[{{Pottasch}(1963)}]{Pottasch_63}
{Pottasch}, S.~R. 1963, \apj, 137, 945, \dodoi{10.1086/147569}

\bibitem[{{Saba}(1995)}]{Saba_1995}
{Saba}, J.~L.~R. 1995, Advances in Space Research, 15, 13

\bibitem[{{Schmelz} {et~al.}(2012){Schmelz}, {Reames}, {von Steiger}, \&
  {Basu}}]{2012ApJ...755...33S}
{Schmelz}, J.~T., {Reames}, D.~V., {von Steiger}, R., \& {Basu}, S. 2012, \apj,
  755, 33, \dodoi{10.1088/0004-637X/755/1/33}

\bibitem[{{Schwab} {et~al.}(2020){Schwab}, {Sewell}, {Woods}, {Caspi}, {Mason},
  \& {Moore}}]{2020ApJ...904...20S}
{Schwab}, B.~D., {Sewell}, R. H.~A., {Woods}, T.~N., {et~al.} 2020, \apj, 904,
  20, \dodoi{10.3847/1538-4357/abba2a}

\bibitem[{{Shanmugam} {et~al.}(2020){Shanmugam}, {Vadawale}, {Patel},
  {Adalaja}, {Mithun}, {Ladiya}, {Goyal}, {Tiwari}, {Singh}, {Kumar},
  {Painkra}, {Acharya}, {Bhardwaj}, {Hait}, {Patinge}, {Kapoor}, {Kumar},
  {Satya}, {Saxena}, \& {Arvind}}]{shanmugam20}
{Shanmugam}, M., {Vadawale}, S.~V., {Patel}, A.~R., {et~al.} 2020, Current
  Science, 118, 45, \dodoi{10.18520/cs/v118/i1/45-52}

\bibitem[{{Sylwester} {et~al.}(2019){Sylwester}, {Sylwester}, {Siarkowski},
  {Phillips}, {Podgorski}, \& {Gryciuk}}]{2019SoPh..294..176S}
{Sylwester}, B., {Sylwester}, J., {Siarkowski}, M., {et~al.} 2019, \solphys,
  294, 176, \dodoi{10.1007/s11207-019-1565-9}

\bibitem[{{Vadawale} {et~al.}(2014){Vadawale}, {Shanmugam}, {Acharya}, {Patel},
  {Goyal}, {Shah}, {Hait}, {Patinge}, \& {Subrahmanyam}}]{2014AdSpR..54.2021V}
{Vadawale}, S.~V., {Shanmugam}, M., {Acharya}, Y.~B., {et~al.} 2014, Advances
  in Space Research, 54, 2021, \dodoi{10.1016/j.asr.2013.06.002}

\bibitem[{{Vanitha} {et~al.}(2020){Vanitha}, {Veeramuthuvel}, {Kalpana}, \&
  {Nagesh}}]{vanitha20}
{Vanitha}, M., {Veeramuthuvel}, P., {Kalpana}, K., \& {Nagesh}, G. 2020, in
  Lunar and Planetary Science Conference, Lunar and Planetary Science
  Conference, 1994

\bibitem[{{Walker}(1972)}]{1972SSRv...13..672W}
{Walker}, A.~B.~C., J. 1972, \ssr, 13, 672, \dodoi{10.1007/BF00213505}

\bibitem[{{Walker} {et~al.}(1974{\natexlab{a}}){Walker}, {Rugge}, \&
  {Weiss}}]{1974ApJ...188..423W}
{Walker}, A.~B.~C., J., {Rugge}, H.~R., \& {Weiss}, K. 1974{\natexlab{a}},
  \apj, 188, 423, \dodoi{10.1086/152730}

\bibitem[{{Walker} {et~al.}(1974{\natexlab{b}}){Walker}, {Rugge}, \&
  {Weiss}}]{1974ApJ...192..169W}
---. 1974{\natexlab{b}}, \apj, 192, 169, \dodoi{10.1086/153048}

\bibitem[{{Warren}(2014)}]{2014ApJ...786L...2W}
{Warren}, H.~P. 2014, \apjl, 786, L2, \dodoi{10.1088/2041-8205/786/1/L2}

\end{thebibliography}
\bibliographystyle{aasjournal}


\end{document}